\documentclass[twocolumn,showpacs,amssymb,aps,nofootinbib,floatfix]{revtex4}

\usepackage{amsmath}
\usepackage{epsfig,color}

\newcommand{\ave}[1]{\left\langle #1 \right\rangle}
\newcommand{\modulo}[1]{\left| {#1} \right|}
\newcommand{\eps}{\varepsilon}

\begin{document}
\title{Bulk viscosity driven
clusterization of quark-gluon plasma and early freeze-out in relativistic
heavy-ion collisions}
\author{Giorgio Torrieri}
\affiliation{Institut f\"ur Theoretische Physik,
  J.W. Goethe Universit\"at, Frankfurt am Main, Germany\\
torrieri@fias.uni-frankfurt.de}
\author{Boris Tom\'a\v{s}ik}
\affiliation{Univerzita Mateja Bela, Bansk\'a Bystrica, Slovakia \\
and Faculty of Nuclear Science and Physics Engineering, Czech Technical University, 
Prague, Czech Republic}
\author{Igor Mishustin}
\affiliation{Frankfurt Institute for Advanced Studies, Frankfurt am Main, Germany \\
and Kurchatov Institute, Russian Research Center, Moscow 123182, Russia.
}

\date{February 17, 2008}
\begin{abstract}

We introduce a new scenario for heavy ion collisions that could solve the lingering problems associated with the so-called HBT puzzle. We postulate that the system starts expansion as the perfect quark-gluon fluid but close to freeze-out it splits into clusters, due to a sharp rise of bulk viscosity in the vicinity of the hadronization transition.   We then argue that the characteristic cluster size is determined by the viscosity coefficient and the expansion rate. Typically it is much smaller and at most weakly dependent of the total system volume (hence reaction energy and multiplicity).  These clusters maintain the pre-existing outward-going flow, as a spray of droplets, but develop no flow of their own, and hadronize by evaporation. We provide an ansatz for converting the hydrodynamic output into clusters.
\end{abstract}
\maketitle

\section{Introduction}
One of the most unexpected, and as yet unexplained, experimental results found at the Relativistic Heavy Ion Collider (RHIC) concerns the description of particle interferometry observables \cite{hbtreview}.   Before RHIC was turned on, it was expected that the deconfined matter would be a highly viscous, weakly interacting quark gluon plasma \cite{danielgyul}.   Thus, ideal hydrodynamics would not provide a good description of flow observables sensitive to the early stages of the collision, such as azimuthal anisotropy \cite{heiselberg} as viscous corrections to these observables would be too large. Nevertheless, after data was released, hydrodynamic simulations offered 
in fact very successful interpretation
of transverse momentum spectra and their azimuthal anisotropy \cite{hydroheinz}. However, hydrodynamics also 
lead to the prediction, that a clear signature for the phase transition would be an increase of the ``out'' to ``side'' emission radius  ratio
(referred to as $R_o$ and $R_s$) due to longer lifetime of the system, caused by the softening of the equation of state in the transition/crossover region \cite{predictions}. This prediction turned out not to be true
\cite{hbtpuzzle}.

Measured parameters $R_{o}$ and $R_{s}$ are  nearly identical.   
Their (positive) difference $R_{o}^2 - R_{s}^2$ is thought to 
correspond---somewhat simplified---to the duration of particle emission.
Hence, it looks like the fireball emits particles almost instantaneously and does not 
show any sign of phase transition or crossover. Hydrodynamics, with ``reasonable'' freeze-out condition (such as a critical temperature of 100 MeV or so) can not describe this even qualitatively.


There could be three possible approaches to the HBT puzzle.  It could be that the system is simply too complicated, 
and that once we include all possible improvements (full 3D calculation, viscosity, hadronic kinetic afterburner, in-medium hadron modifications etc.), everything will fit.
It could also be that we are drastically misunderstanding the data, and the HBT puzzle is a 
symptom of inapplicability of hydrodynamics to heavy ion collisions.
Finally, it could be that the hydrodynamic approach is basically correct, but there is just one element 
of physics relevant to freeze-out that is fundamentally misunderstood.   

The second possibility is unlikely because, in some ways, hydrodynamic prescription {\em does} fit HBT data.   
Scaling of HBT radii with the multiplicity rapidity density $(dN/dy)^{1/3}$, over a large range 
of energies \cite{lisa} is typical for an isentropically expanding fluid that suddenly breaks apart.
The very good description, within parameters compatible with what is needed to describe flow, of the {\em azimuthal} dependence of HBT radii \cite{kolbangle}, also suggests that the hydrodynamic framework is a good ansatz for describing the matter produced in heavy ion collisions {\em up to freeze-out}.

The first possibility appears, however, also problematic:
successful models and/or parametrisations of the freeze-out which describe HBT radii are found in the literature 
\cite{florkowski,sinyukov,budalund,seattle,blastwave}, and they could provide a way to gain insights into what is missing. However, we feel that  successful description involves a dynamical description from initial conditions {\em plus a freeze-out criterion}, rather than a fit to data with assumptions put in ``by hand''.   Such a description is so far lacking\footnote{Some kinetic models incorporating partonic interactions, such as \cite{ampt}, manage to reproduce HBT data for certain values of the parton scattering cross-section.  The interpretation of these results within a collective picture is however not yet fully understood.}.
Furthermore, the most plausible refinements to hydrodynamics, namely
implementation of fully three-dimensional models \cite{hirano} and
the addition of a kinetic theory afterburner \cite{shuryak} do not do anything
to solve the HBT discrepancy, but in fact make it worse, suggesting that the problem is not refinements but rather one large missed physical effect.

One such effect discussed so far in the literature within a hydrodynamic 
context is spinodal clustering \cite{mish1,mish2,randrup,kapusta} driven 
by a first order phase transition (some authors talk about explosive 
freeze-out, within the same context \cite{sudden1,sudden2,sudden3,sudden4}).

While clustering has been applied to describe HBT data \cite{zhang}, such a description is yet to be made fully kinetic.  One reason why such an ansatz has not been accepted so far is that lattice strongly indicates that the 
transition at RHIC energies is not of first-order, but rather a smooth cross-over, and the critical point 
appears at a considerable chemical potential.
Considering the rather universal scaling \cite{lisa,caines,busza} found in soft observables, an alternative which is not dependent on the assumption of first order phase transition is desirable.

In this paper, we propose such a possible alternative freeze-out mechanism, where the dynamics 
explain why it can lead to freeze-out considerably different from the usually
used typical energy density. We argue that it can, in fact, be the basis of
reconciling hydrodynamics and interferometry.   Instead of freeze-out happening at
a critical temperature or energy density, we speculate that the system breaks up into fragments, 
as a result of the bulk viscosity sharply rising close to the phase transition temperature.   
This explanation has the virtue that it is connected to theoretical features of QGP, namely
its near-perfect conformal invariance at high (perturbative)
temperatures, and the existence of a conformal anomaly in the
non-perturbative regime.



\section{The behaviour of bulk viscosity near $T_c$}

The bulk viscosity of high temperature strongly interacting matter has recently been calculated using perturbative QCD \cite{amybulk}, and found to be negligible, both in comparison to shear
viscosity and w.r.t. its effect on any reasonable collective evolution of the system.
This is not surprising:  The QCD Lagrangian, as long as no ``heavy'' quarks are present, is nearly conformally invariant \cite{amybulk}.    Since, within a
fluid, the violation of conformal symmetry is linearly proportional to
a bulk viscosity term \cite{lifshitzlandau}, the near conformal invariance of the QCD
Lagrangian should guarantee that bulk viscosity is nearly zero,
\textit{in the perturbative regime}.

In the hadron gas phase, of course, the numerous scales associated
with hadrons render conformal invariance a bad symmetry, and hence it
is natural to expect that bulk viscosity is not negligible. 

This is, again, rooted in a fundamental feature of QCD:  the
non-perturbative \textit{conformal anomaly}, that manifests itself in the
scale (usually called $\Lambda_{QCD}$) at which the QCD coupling
constant stops being small enough for the perturbative expansion to
make sense.  This scale coincides with the scale at which confining forces hold hadrons together.

This violation of conformal invariance is not seen perturbatively, but
should dominate over the perturbatively calculated bulk viscosity as temperature drops
close enough to the QCD phase transition.

What happens to \textit{bulk} viscosity in this regime, where hadrons are not yet
formed, presumably the matter is still deconfined, but conformal
symmetry is badly broken?  While we can not as yet calculate this rigorously, 
there are compelling arguments \cite{pratt,kharbulk,kkbulk} 
that bulk viscosity rises sharply, or even diverges, close to the phase transition temperature.

Lattice simulations find that $T^{\mu}_{\mu}$ (=0 for a
conformally invariant system), increases  rapidly close to
$T_c$.  Remembering that the shear ($\eta$) and bulk ($\zeta$) viscosities
roughly scale as \cite{hosoya,jeonvisc,weinberg}
\begin{eqnarray}
\label{bulkgeneral}
\eta & \sim & \tau_{\rm elastic} T^4 \\
\zeta & \sim & \left( \frac{1}{3} - v_s \right)^2 \tau_{\rm inelastic} T^4
\end{eqnarray}
where $\tau_{\rm (ine)elastic}$ refers to the equilibration timescale of (ine)elastic
collisions.
Assuming  $\tau_{\rm inelastic} \sim 1/T$ allows to extract the bulk viscosity from the lattice, and yields a sharp rise close to $T_c$.   This can be more formally seen from finite temperature sum rules in conjunction with lattice data \cite{kharbulk,kkbulk}.

The rise is, in fact, likely to be considerably sharper than \cite{kharbulk} suggests.    The dependence of $\tau_{\rm inelastic}$ on temperature can be guessed from
the fact that, at $T_c$, the quark condensate $\ave{q \overline{q}}$ 
acquires a finite value, and the gluon condensate
$\ave{G_{\mu \nu} G^{\mu \nu}}$ sharply increases at the phase
transition.   ``Kinetically'', therefore, timescales of processes
that create extra $q \overline{q}$ and $G G$ pairs should diverge
close to the phase transition temperature, by analogy with the divergence of the spin correlation length in the Ising model close to the phase transition. 
Numerical studies with viscous hydrodynamics and a chiral model \cite{pratt} seem
to confirm that the second-order chiral phase transition makes the
viscosity diverge. 

The sharp rise of bulk viscosity can also be understood within string kinetics:
confinement, microscopically, can be thought of
as a ``string tension'' appearing in the potential.  Given the small
mass of light quarks (and hence ease of creating $q \overline{q}$
pairs), the appearance of  even small string tension will lead to a
small ``preferred scale'' at which strings break.   Hence, conformal
symmetry should be quickly badly violated right at the deconfinement
phase transition.
In particular, in a regime where
the momentum exchange of the average collision is more than enough to
break the string, the relevant degrees of freedom are still quarks, not mesons, and the shear viscosity is still low, 
a profound change happens:  each previously elastic collision, that
before just diffused momentum, becomes inelastic, where the final state has
less kinetic energy than the initial state.
Even if this difference (the energy needed to break the string) is low, over many collisions, the heat energy would
be converted into creating more slightly colder, less pressing particles.
That's exactly the kind of processes that contribute most to bulk viscosity \cite{jeonvisc}.

These arguments give evidence to the conjecture that, close (from
above) to $T_c$, bulk viscosity goes rapidly from a
negligible value to a value capable of \textit{dominating} the
collective
evolution of the system.
That this transition is sharp can be seen by the sharpness of the lattice deconfinement transit
from lattice flavor correlations studies (such as $\ave{\Delta B \Delta S}$ \cite{bs})
seem to confirm that, immediately above $T_c$, the relevant degrees of freedom
become quasi-particles similar to the asymptotically free quarks.
It is therefore likely that non-perturbative effects 
(such as the conformal anomaly) go away soon above $T_c$.
Conversely, they should appear quite suddenly if $T$
approaches $T_c$ from above, in an expanding cooling fluid.

In the next section, we will show how this picture could yield 
a freeze-out scenario that has the potential to resolve the HBT puzzle.


\section{Clustering at the viscosity peak}
\label{oil2honey}

\begin{figure*}[tb]
\centerline{\epsfig{file=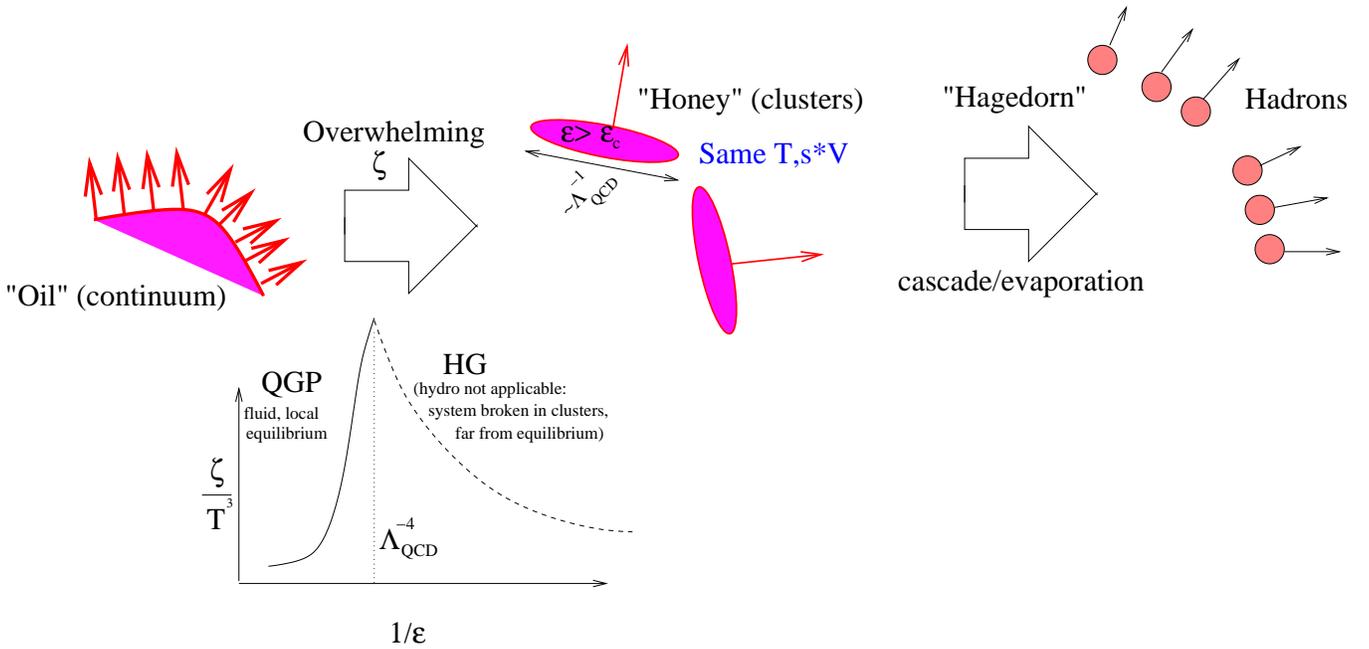,width=18cm}}
\caption{\label{diagram} (Color online) 
Fragmentation of the fireball due to sharply increasing bulk viscosity as
the temperature decreases. Matter which expanded easily before we describe as
oil. It suddenly becomes very rigid against expansion (described as honey in the 
figure) and breaks up into fragments. Hadrons evaporate from these fragments. 
}
\end{figure*}

It has been noted that bulk viscosity could be helpful in making
$R_{o}$ and $R_{s}$ agree with experiment \cite{muronga}.   Of
course, too large bulk viscosity, in the context of heavy ion
collisions, would just mean that the approximations on which hydrodynamics is based are
not accurate (the mean free path is {\em not} negligible w.r.t.\ the
system's inhomogeneities) and an approach not relying on assumed near-equilibrium might be
necessary \cite{pratt,kapusta}.  

One guess for the qualitative behaviour of such a system is
illustrated in Fig.~\ref{diagram}:  one just
needs to think what would happen if a rapidly expanding compressible fluid suddenly becomes rather rigid, sticky and resistant to further expansion and deformation. \footnote{It should be noted that QCD matter is different enough to ``ordinary'' matter that concepts such as ``solid'',''liquid'',''sticky'' etc. are misleading, since these words carry tacit assumptions that are valid in the ''everyday'' world but potentially badly broken in QCD.  In solid state physics, short-range potentials are usually dominated by steep ``walls'' driven by the Pauli exclusion principle.  Thus, materials with small inter-molecular distances, either small-viscosity ``good liquids'' or large viscosity ``solids'', are almost always highly incompressible, because of the steep inter-molecular potential at small average molecular separation.
This incompressibility is usually assumed in the definition of both ``liquid'' or ``solid''.  In QCD there is no such short-range repulsive potential, the short-range EoS is the conformally invariant ideal gas one, and lattice calculations show that compressibility of QCD matter is never high even when the bulk viscosity rises close to $T_c$.   The highly viscous phase is thus not a ``solid'' or a ``glass'', since these materials are usually defined as incompressible, while highly viscous QCD matter can be compressed easily {\em if done infinitely slowly}.  At finite compression/expansion speed, on the other hand, the system becomes ``solid-like'' due to the viscosity, an argument central to our subsequent reasoning.  Such characteristics are, to our knowledge, highly unusual in solid state physics yet well motivated in QCD}
If the material spent a considerable amount of time in the low viscosity 
phase there will be pre-existing collective flow that is pushing the
system outwards.    The inhomogeneities of this flow will rapidly generate
strong viscous forces, which will tend to decelerate and stop the expansion. 
These forces, by causality, will not be able to quickly overcome the
pre-existing flow {\em globally}, but more than enough to overcome it {\em locally}.
If the appearance of viscosity is sharp enough, these forces cannot overcome the inertia 
of the system and 
it is natural to suppose that the system will be
rapidly broken apart into small fragments, each flowing with pre-existing flow, with roughly QGP density.

It should be clearly stated that although this scenario shares some superficial similarities 
with the more usual nucleation picture analyzed, e.g., in \cite{kapusta,sudden2,mish1,mish2}, it is 
physically completely different.    

In \cite{kapusta}, hadronic bubbles form in a steady supercooled QGP medium, and the role of viscosity is to dissipate the latent heat during bubble growth.   The nucleation examined here, on the other hand, occurs due to the interplay of a suddendly appearing viscosity with the pre-existing advective forces.

The scenario in \cite{kapusta} requires a first-order phase transition, our does not. 
The nucleation in \cite{kapusta} proceeds via creation of critical bubbles
whose size is determined only by thermodynamical quantities, viscosity influences the dynamics of bubble growth but not the critical size, and global collective expansion plays no role. 
We require a robust pre-existing global expansion, and a sharp peak of viscosity
at the critical temperature that forces the system to disintegrate into fragments which, in principle, can depend on both local and global system properties.
 
 The bubbles in \cite{kapusta} are made of hadron gas. Ours are evaporating droplets of hot QGP.      In \cite{kapusta}, clustering will entail an entropy increase, proportional to the latent heat and explained by the different entropy density of the two phases.   In our approach, the formation of clusters should quickly kill off $\partial_{\mu} u^\mu$, so entropy generation ($\sim (\partial_\mu u^\mu)^2$) during clustering should be negligible.

Thus, this work and pre-existing clustering \cite{kapusta} should not be treated as complementary
descriptions of the same phenomena, but as competing scenarios to be 
differentiated at the theoretical level (are the conditions for either scenario
relevant to heavy ion collisions?) or through experimental data (since our clusters are very different from those examined
in \cite{kapusta,sudden2,mish1,mish2}).

Relativistically, too high a viscosity indicates that the system is too far from equilibrium for the Navier-Stokes equations to be a good description.  At best, higher order corrections to hydrodynamics, e.g. \cite{israelstewart} become dominant.  At worst (and more likely), the whole expansion in flow gradients becomes divergent.   The scenario considered here represents a {\em guess} of how the system could evolve after hydrodynamics breaks down as an appropriate physical description.  

While this guess, with reasonable timescales/cluster sizes, should not suffer from the causality violation
pathologies that affect first order viscous hydrodynamics, we do not at the moment see a way to formally assess its likelihood from transport theory arguments beyond
deriving some quantitative conditions for this scenario to be plausible. We shall do so in the rest of this section.

The first condition is that forces due to bulk viscosity must overwhelm advective forces
\begin{equation}
\alpha_1 = \frac{p}{ \zeta \partial_\mu u^\mu} \ll 1
\end{equation}
where $p$ is the pressure and $u^\mu$ velocity field.
If bulk viscosity {\em diverges}, as argued in this work and \cite{kharbulk}, the applicability of this condition is assured. While an ab initio lattice extraction involving pure gluons \cite{bulklattice} does not have evidence for
a divergence, they estimate
\begin{equation}
 0.5<\frac{\zeta}{s} (T=1.02 T_c) < 2
\end{equation}
($s$ is entropy density.)
Assuming an inviscid conformal Bjorken \cite{bjorken} dynamics until the rise in bulk viscosity, we have
\begin{eqnarray}
\partial_\mu u^\mu = \tau^{-1} \\
\frac{T}{T_0} = \left( \frac{\tau_0}{\tau}  \right)^{1/3}\\
\frac{\varepsilon}{\varepsilon_0} = \left( \frac{\tau_0}{\tau}  \right)^{4/3}
\end{eqnarray}
where $\varepsilon$ stands for energy density.

We can then obtain an estimate for $\alpha_1$ in terms of the initial temperature $T_0$ and thermalization timescale $\tau_0$ 
\begin{equation}
\alpha_1 \sim \frac{s}{\zeta} \frac{T_c \tau_c}{4} \sim \frac{s}{\zeta} \frac{T_0^3 \tau_0}{4 T_c^2}
\label{condition1}
\end{equation}
where $\tau_c =\tau (T=T_c)$
This might be less than unity even in the ``worst case scenario'', although, of course, the divergence of $\zeta$ would make clustering much more plausible.

The second condition is that the appearance of the viscosity divergence is {\em sudden} enough for it not to be dissipated by hydrodynamic evolution.     For a qualitative estimate,
\begin{equation}
\frac{1}{\zeta} \frac{d \zeta}{d \tau} \gg \partial_{\mu} u^{\mu} \Rightarrow \frac{1}{\zeta} \frac{d \zeta}{d T} \gg \frac{d \tau}{\tau d T} 
\end{equation}
or the onset of bulk viscosity will just be dissipated through 
hydrodynamic evolution.   Once again, a sharp divergence of $\zeta$ will ensure that this condition is satisfied.  Considering, as a toy model, a Gaussian peak of the evolution of $\zeta$ w.r.t. temperature, and assuming $\sigma_\zeta$ to be the width of the peak, we get
\begin{equation}
\frac{1}{\zeta} \frac{d \zeta}{d T} = \frac{2 (T - T_c)}{2 \pi \sigma_\zeta^2} \gg \frac{1}{\tau} \frac{d \tau}{d T}
\end{equation}
fitting a $\sigma_\zeta$ to the output of \cite{kharbulk} and comparing with a Bjorken estimate for $dT/d \tau$ should convince us that this criterion is very plausible as $T$ approaches $T_c$ from above.

The third condition is that terms in the second order of the flow gradient are {\em not} relevant for the system under consideration.  The effect of these is the emergence of a time-scale (the relaxation time, $\tau_\Pi$), which delays the appearence of viscous forces from the built-up of the flow gradient.
If the divergence in $\zeta$ is too sharp around $T_c$, or if the relaxation time is too long, the singularity in $\zeta$ will have no effect on the dynamics:
by the time the viscous forces turn on, the system has allready been cooled
to below $T_c$ and viscosity is not anymore singular.
If $\Delta T$ is the width of the peak in $\zeta$, this condition can be quantitatively estimated as
\begin{equation}
\alpha_2 =\frac{\tau_\Pi}{\Delta T} \frac{dT}{d \tau} \ll 1
\label{condition2}
\end{equation}
$\tau_\Pi$ is famously difficoult to estimate from quantum field theory.
The estimate closest to strongly coupled QCD we have is provided by calculations in super-symmetric Yang-Mills theories\footnote{Note that this estimate is for a {\em conformally invariant theory}, so it is in direct contradiction with our scenario, and hence particularly unreliable. Still, nothing more realistic is available at the moment}. \cite{janik}
\begin{equation}
\tau_\Pi = \frac{1 - \log 2}{6 \pi T}
\label{taupi}
\end{equation}
Estimating from \cite{kharbulk} $\Delta T \sim 0.1 T_c$, and assuming once again a Bjorken-type evolution before the divergence, we find 
\begin{equation}
\alpha_2 \sim  \frac{10(1 - \log 2)}{18 \pi}  \frac{1}{T_c \tau_c}
\end{equation}
since in general, $\alpha_2 \ll \alpha_1$ the fulfillment of the second condition is more likely than the first at all energies.
The introduction of multi-dimensional expansion (rather than the 1d case of \cite{bjorken}) will for sure increase $\partial_\mu u^{\mu}$ and $dT/d\tau$,lowering $\alpha_1$ and raising $\alpha_2$ (Since $\alpha_2 \ll \alpha_1$, this raises the possibility for the presently considered scenario).  

The estimates here, given our very limited understanding of some key parameters, should only be taken to understand that clustering is {\em not} outright excluded.   Our aim in the subsequent sections of the paper is to try to quantitatively estimate some phenomenological consequences of this scenario, and try to connect it to experimental data.
\section{An estimate for the cluster size}
The simplest argument is based on the assumption that bulk viscosity diverges at the 
critical point and therefore decouples from the problem. The relevant scales are set 
by $\Lambda_{QCD}$ and $T_c$. 
Each fragment will have a typical size
$R_c \sim \Lambda_{QCD}^{-1}$ (the preferred scale of the system) 
and move in the direction
determined by its pre-existing collective flow field.  
The typical energy density in fragments is  about $aT_c^4$ with  $a=\pi^2\nu_{QGP}/30$, where $\nu_{QGP}\approx 30$ is effective number of degrees of freedom in the quark-gluon plasma (QGP). For our rough estimates we take the critical temperature $T_c=165$~MeV and $\Lambda_{QCD}=250$~MeV. Then the typical fragment mass is estimated as
\begin{equation}
\label{cluster}
M\propto \frac{4}{3} \pi R_c^3\frac{\pi^2}{30}\nu_{QGP}T_c^4\approx 2\, 
\mbox{GeV}.
\end{equation}
Such a cluster (droplet) will decay into about 10 pions or a few heavier hadrons.
%
Note that this estimate is good for a cluster containing no
strangeness or baryon number.   To handle these, Eq.~\eqref{cluster} needs 
to be updated to accommodate strangeness and
baryon content, perhaps using the methods outlined in \cite{majumder}.
Naively, the higher energy content of baryon and strangeness rich QGP
should increase $M_{\rm cluster}$, so that clusters containing baryons
(strange and non-strange) should also have the mass of high-lying
baryonic resonances and decay into several particles.
Note also that $T$ might well be considerably larger than the phase
transition (let alone the chemical freeze-out) temperature.  It is
simply the temperature at which the bulk viscosity starts becoming
strong enough to locally counteract the built-up flow.
The large bulk viscosity, collective manifestation of the inter-particle confining
potential, will prevent these fragments from expanding further.
They should therefore be considered Hagedorn-style ``fireballs''
rather than as expanding fluid clumps.    Cascading of these fireballs into
the ground-state hadrons produces the hadrons at ``chemical
freeze-out''.

The fact that the scale suggested here is similar to the hadronic scale 
begs the question of whether this picture is significantly different from the 
``usual'' Cooper-Frye particle emission
picture.  The difference is that within the Cooper-Frye scenario, the mostly produced particle  is the ``massless'' pion, while in our scenario only systems having ``hadronic'' $\sim \Lambda_{QCD}$ mass scale are created at hadronization.  These systems, furthermore, are not ``particles'' (zero temperature states) but rather finite temperature fireballs, although it is reasonable for them to transform into Hagedorn-type resonances and decay.

In the presence of collective expansion $\Lambda_{QCD}$ might interplay with other scales of the problem set by 
expansion velocity gradients. Let us use them in an estimate of the size of fragments
related to the dynamics of the expansion. 
In order to do so it is useful to recall that 
the energy momentum tensor, with vanishing shear viscosity but non-vanishing
bulk viscosity is
\begin{equation}
T^{\mu\nu} = (\eps + p) u^\mu u^\nu - p g^{\mu\nu} + \zeta\, \partial_\rho u^\rho \, (g^{\mu\nu} - u^\mu u^\nu)
\end{equation}
From energy-momentum conservation $\partial_\mu T^{\mu\nu} = 0$ we then obtain the rate of energy density
decrease
\begin{equation}
\label{erate}
\frac{1}{\eps}u^\mu \partial_\mu \eps = 
\frac{\eps+ p - \zeta \partial_\rho u^\rho}{\eps}  \partial_\mu u^\mu\, .
\end{equation}
Note that when $\zeta \partial_\rho u^\rho \sim p$ the energy density decreases at the same rate as 
if no work was performed in case with vanishing viscosity. For lower rates of the energy density decrease
the expansion even {\em decelerates}. Microscopically, this is mediated by inter-particle forces which
hold the system together. It can happen that the inertia of the bulk overcomes these forces and the 
system thus {\em fragments}. 

In order to obtain a more quantitative estimate of droplet size, we determine it
by the balance of deposited energy and collective expansion energy.
According to the definition of viscosity, it determines the amount of energy deposited per 
unit volume and unit time, i.e.
\begin{equation}
E_{\rm dis}=\int dV \int d\tau \zeta(\partial_{\mu} u^{\mu})^2,
\end{equation}
where $\zeta$ is bulk viscosity and $u^\mu$ collective 4-velocity. For simplicity let us assume again the Bjorken \cite{bjorken} picture. 
Then $\partial_\mu u^\mu=1/\tau$ and the 3-velocity is $v_z=z/t$. If bulk viscosity is indeed rapidly divergent at 
$T_c$, we can replace it with the $\delta$-function
\begin{equation}
\zeta(\tau)= 
\zeta_c T_c \delta \left( T(\tau) - T_c \right) = 
\zeta_c T_c \left. \frac{d \tau}{d T } \right|_{T=T_c} \delta(\tau-\tau'_c),
\end{equation}
where
$\zeta_c$ is a model parameter which should be given by deeper theoretical consideration.  If we call $\tau'_c =  T_c \left. \frac{d \tau}{d T } \right|_{T=T_c}$  we get
\begin{equation}
E_{\rm dis}=SL\frac{\zeta_c}{\tau'_c},
\end{equation}
where $S$ is the transverse area of the Bjorken cylinder and $L$ is the droplet longitudinal size. We consider a 
droplet whose center of mass is located at $z=0$ (though this assumption is not really important due to the 
boost invariance of the system).

The kinetic energy of droplet's expansion, which is in fact dissipated due to viscosity, is in non-relativistic limit
\begin{equation}
E_{\rm kin}=\frac{1}{2}\int dV\, \eps(\tau)v_z^2,
\end{equation}
where $\eps(\tau)$ is the  internal energy density of the fluid. It is of course a function of time but the above expression contains only volume integration. Let us evaluate the integral at the critical point, when actual break-up happens, then
\begin{equation}
E_{\rm kin}=\frac{S\, \eps_c}{24t_c^2}L^3.
\end{equation} Taking $t_c\approx\tau'_c$, we get finally
\begin{equation}
\label{dsize}
L^2=\frac{24\zeta_c\tau'_c}{\eps_c}\, .
\end{equation}
Notice that $\tau'_c$ in the numerator is actually the inverse expansion rate 
$\partial^\mu u_\mu$. Thus the droplet size squared is inversely proportional to 
the expansion rate. 
Within this scenario the droplet size will
grow with the lifetime of the hydrodynamic stage (from the initial temperature $T_0$ to $T_c$), but the growth will 
generally be 
slower than linear.
For our toy model example where the system has a conformal equation of state
 and Boost-invariance ($dN/dy \sim \epsilon_0^{3/4} \sim \tau'_c$), this growth will be $\sim (dN/dy)^{1/2}$, but it 
is likely to be slower than that when  transverse expansion is considered.

Whether the cluster size is indeed only dependent on the internal scale of the system $\Lambda_{QCD}$ (Eq. \ref{cluster}) or on an interplay between the internal and collective scales (Eq. \ref{dsize}) is difficult to determine from first principles, as it depends on a quantitative understanding of the details of the non-equilibrium evolution around $T_c$.

The main point argued in the last section, one that does not depend on these
details, is that the sharp rise of bulk viscosity could force the system to break up into disconnected fragments, of a scale and lifetime much smaller than
the size of the system ($\mathcal{O}(1\phantom{A} \mathrm{ GeV}))$.  
These 
clusters 
then flow 
apart with 
pre-existing flow velocity and, presumably, decay by Hagedorn cascading.   In the next three sections we shall examine the effect this kind of freeze-out has on heavy ion phenomenology.



\section{Phenomenology of clustering}

While the HBT puzzle is our main experimental motivation for introducing a
qualitatively new freeze-out scenario, several
observables, aside from particle interferometry, could imply clustering.  
In this section, we point out a list of such phenomena.  In each of these, the evidence for clustering is by no means overwhelming, and alternative explanations for each of these phenomena exist.   Nevertheless, it is worthwhile to point these phenomena out individually as candidates for contact between the model presented here and experimental data. 
\begin{itemize}
\item As pointed out in \cite{etele}, a highly viscous but hydrodynamic evolution is constrained experimentally by multiplicity measurements.
The dependance of multiplicity on centrality has been shown to be well
described through exclusively initial conditions (Glauber model, or, at high energies, the Color Glass Condensate).  Since expansion of a viscous fluid generates entropy at the rate \cite{lifshitzlandau}
\begin{equation}
 \partial_\mu s^\mu \sim  \zeta \left( \partial_\mu u^\mu
\right)^2 
\end{equation}
too much viscosity at any stage during the hydrodynamic evolution would spoil the agreement between experimentally observed multiplicity and ansatze based on initial conditions.    This is a potential problem of {\em all} attempts of solving the HBT puzzle through viscous but hydrodynamic evolution \cite{tsunami,muronga}

Viscosity-driven clustering would not have this problem, since within the cluster
all relative motion is very quickly killed.  Thus, while $\zeta$ might diverge, $\partial_\mu u^\mu$ would vanish. Since entropy production rate
is quadratically proportional to the latter, we would expect the entropy
content of the system to not be significantly changed during the
clustering and freeze-out phase.

\item The very fact that a ``single freeze-out model'' \cite{florkowski,ourfreeze} works much better than naively expected in describing soft observables in heavy ion collisions suggest that ``something'' is decreasing hadronic interactions after chemical freeze-out below the expected rate.   Clustering of the system into smaller sub-systems that decay after a finite time, during which they flow out with the pre-existing flow, would have just such an effect.

\item The {\em over-}abundance of certain resonances ($\Xi^*$, $\Delta$, $\Sigma^*$) with respect to 
even chemical freeze-out expectations \cite{salur,witt}  could also be nicely explained in terms of clustering.   Clusters can be considered as highly excited Hagedorn tower resonances.  It is therefore natural to suppose that they could decay through a cascade  
down the Hagedorn ``tower'', and hence through the production of resonances.
Hence,  ratios such as $\Sigma^*/\Lambda$ and $\Xi^*/\Xi$ would be correspondingly enhanced.
Stable particles should still be well described by the
statistical model: the final hadron abundance will be a collection of a large
number of microcanonically decaying fireballs, carrying 
grand-canonically distributed energy and quantum numbers. 

\item The scaling of $p_T$ fluctuations provides direct evidence
that  particles are emitted from clusters, containing a small ($\sim 5$)
number of particles independently of collision energy or centrality \cite{ptfluct}.  The under-prediction, by the equilibrium statistical model, of fluctuations of ratios such as $K/\pi$ \cite{sqm2006} compounds this evidence, since cluster emission would enhance fluctuations of multiplicity yields and ratios.
The forward-backward multiplicity correlations \cite{fbcorrel} and angular correlations in 
Cu+Cu collisions  at RHIC \cite{roland_darmstadt} also indicate the presence of clusters. 

\item Clustering into fragments of size about 1~GeV could provide an explanation for the invariant mass systematics of the inverse slopes $T_{\rm eff}$ observed at SPS and RHIC energies, both for stable particles (Fig.~1 of \cite{slopes_stable}) and electromagnetic resonance decays ($\rho \rightarrow \mu^+ \mu^-$, 
Fig.~1 left panel of \cite{slopes_reso}).
All inverse slopes for particles less massive than roughly 1~GeV seem to rise with mass, as expected, approximately, for ``blue-shifted'' thermal emission where the inverse slope combines temperature and flow 
($T_{\rm eff} \simeq T + \ave{v_T} M$).   For masses larger than 1 GeV, however, $T_{\rm eff}$ 
is nearly independent of mass.
Clustering, if all clusters have a mass of about 1~GeV, could provide a natural explanation for this observation:  Below the invariant mass of 1~GeV, particles are predominantly emitted from a single cluster, and hence maintain the memory of that cluster's flow.   Above a mass of 1~GeV, however, particles have to be emitted either before cluster formation, or through cluster fusion, and hence there is occasion for the flow to be ``forgotten''.
\end{itemize}


\section{Looking for clusters in HBT \label{hbt_clusters}}

We start by  noting \cite{hydroheinz,wiedemann} that, in
the out-side-long coordinate system, HBT radii are directly related to
the system's spacetime correlation tensor\footnote{Here $l$ (``long'') is the z direction (parallel to the
beam), $o$ (``out'') is the direction of the pair momentum, and $s$ (``side'') is the
cross product of the previous two.}
\begin{eqnarray}
R_s^2(K) &=& \ave{(\Delta x_s)^2} \label{rside}\\
R_o^2(K) &=& \ave{(\Delta x_o)^2} - 2 \frac{k_T}{k_0} \ave{\Delta x_o
  \Delta t} \nonumber \\
  && \qquad \qquad \qquad + \left(  \frac{k_T}{k_0} \right)^2  \ave{(\Delta t)^2} \label{rout}
\end{eqnarray}
and, for pairs of particles having zero net longitudinal momentum
\begin{eqnarray}
R_l^2(K) &=& \ave{(\Delta x_l)^2} \label{rl}
\end{eqnarray}
where the $k$ vector is the sum of the two momenta (the first element, $k_0$, is $\simeq \sqrt{m^2 + \vec k^2}$).
Averaging is done using the emission function
\begin{equation}
\ave{A}(K) = \frac{\int A(x) S(x,K) d^4 x}{\int S(x,K) d^4 x} \, .
\end{equation}
$R_l$ is straight-forwardly related to the longitudinal length of the
fireball.   A correct treatment of deviations from boost-invariance
should therefore also contribute to an improvement of  current discrepancy  
with experimental data\footnote{Please refer to
\cite{hydroheinz}, on the current status of hydro-experiment HBT comparisons}.

As remarked in \cite{hydroheinz}, the $R_o
\sim R_s$ result is not easy to reconcile with naive hydrodynamics
plus a straight-forward (critical temperature) emission because:
\begin{itemize}
\item  The higher the energy, the longer the emission  time,
the larger is the expected discrepancy between $R_o$ and $R_s$.  If the
system starts close to the mixed phase, the timescale of freezing out
should be longer still due to the softest point in the equation of state.
Hence, a generic prediction from Eqs. \eqref{rside} and \eqref{rout} is that 
$R_o/R_s>1$, broadly increases with energy, and exhibits a peak when the energy 
density is such that the system starts within, or slightly above the mixed phase.  This is in direct contrast with
experimental data, where $R_o/R_s \simeq 1$ is a feature at all reaction
energies.
\item Generally in a hydrodynamic model  the
$\ave{\Delta x \Delta t}$ correlation is negative, since particles on
the outer side are the first to freeze-out.   This increases
$R_o/R_s$ further (cf.\ eq~\eqref{rout}).  
Time dilation due to transverse flow does not help
enough, as calculations show.   
\end{itemize}
It is immediately apparent that clustering can help solving both of
these problems.
\begin{itemize}
\item Cluster size, density and decay timescale, is approximately independent of either
reaction energy or centrality, as can be deduced from Eq.~\eqref{dsize}. 
Hence, the near
energy independence of the (comparatively short) emission timescale,
and hence of $R_{o}/R_{s}$, should be recovered.
\item If the decay products do not interact (or do not interact much)
after cluster decay, it can also be seen that $\ave{\Delta x \Delta t}$
can indeed be positive:  outward clusters are moving faster,
resulting in time dilation.   
This effect can be offset by time dilation of
cluster decay by increasing the temperature at which clusters form,
or by increasing cluster size.
\end{itemize}
Recovering the  linear scaling of the radii with $(dN/dy)^{1/3} (\sim
N_{\rm clusters})$  \cite{lisa}, 
while maintaining the correct $R_{o}/R_{s}$ is also possible if
the clusters decay when their distance w.r.t.\ each other is
still comparable
to their intrinsic size.

Quantitative  calculations are necessary before determining whether these
constraints can be satisfied.  The technical details of how to perform such calculations, 
from a hydrodynamic code output with a critical temperature and cluster size, are outlined in the Appendix.  
Hydrodynamics output is needed to specify the cluster flow array $u^\mu_i$ and emission array $\Sigma_\mu^i$, (defined in Eq. \ref{clusterflow}).   

The bulk-viscosity-driven freeze-out adds another parameter to ab initio HBT calculations:  
in addition to critical temperature/energy density, we now have the cluster size.
To see whether this helps solving the HBT problem, output from hydrodynamics with a high ($T \sim T_c$) freeze-out temperature should be fragmented into clusters with a certain distribution in size, 
which then produce hadrons according to the prescription in the Appendix.

If this ansatz, and a reasonable mean/variance do reproduce the observed $R_{o},R_{s}$ from a realistic 
hydrodynamics output, it would provide a strong motivation for looking for clusters in event-by-event physics.
Cluster-driven symmetry breaking should also lead to distinctive 
signatures in the multipole expansion of the correlation function 
\cite{daniel}

%
%

\section{Discussion and conclusions}

We have described a mechanism to generate fragments that is solidly grounded in QCD, and does not require a first order phase transition.
Hence, it is possible that hadronization is governed by this mechanism in all regimes where an approximately locally thermalized deconfined system is produced.

Potentially, this mechanism can solve the HBT problem by adding a further ``free parameter'' to the system: 
the cluster size.
Using the methods described in Section \ref{hbt_clusters}, it is possible to see whether a given cluster distribution, matched to the hydrodynamic output with the freeze-out criterion tuned to cluster formation, could reproduce the measured HBT radii.


Future work in this direction includes both a quantitative comparison between HBT data and the model (with a proper hydro input), as well as signatures for clustering in event-by-event physics.


\acknowledgments
GT would like to thank the Alexander von Humboldt Foundation and Frankfurt
University for the support provided, and to Mike Lisa, Sangyong Jeon, Guy Moore and Johann Rafelski for fruitful discussions. 
BT acknowledges support from 
VEGA 1/4012/07 (Slovakia) as well as MSM 6840770039 and LC 07048
(Czech Republic).
IM acknowledges support provided by the DFG grant 436RUS 113/711/0-2
(Germany) and grants RFFR-05-02-04013 and NS-8756.2006.2 (Russia).

\appendix
\section{Emission function for clusters}
The HBT emission function of a fluid breaking up into identical clusters which
then decay should be given by a sum of cluster emission functions
\begin{equation}
\label{clustersum}
S(x,p) = \sum_{i} S_i (x-x_0^i,p) 
\end{equation}
At the cluster rest frame we suppose, in accordance with the  ansatz described in Section~\ref{oil2honey}, 
that $S_i$ is a simple Gaussian with no further
structure or flow.  This cluster decays, also via a Gaussian function after a time $\tau \sim
\Lambda_{QCD}^{-1}$ after formation, a scale also similar to it's radius.
Normalizing to the number of particles per cluster, and in the
Boltzmann approximation 
\begin{equation}
\label{comovingsource}
S_i (x',p') = \frac{1}{(2\pi)^3}\, \frac{1}{\tau} e^{-E'/T}
e^{-(t'^2+x'^2+y'^2+z'^2)/(2 \tau^2)}
\end{equation}
Collective (cluster) velocity and the hypersurface on which clusters 
are generated are expressed as
\begin{equation}
u^{\mu}_i = \left( \begin{array}{c} \cosh y_{Li} \cosh y_{Ti} \\
\sinh y_{Ti} \cos \theta_i \\
\sinh y_{Ti} \sin \theta_i \\
\sinh y_{Li} \cosh y_{Ti} 
\end{array} \right)
\quad
\Sigma^{\mu}_i = \left( \begin{array}{c}t_{fi} \cosh y_{Li} \\
r_{fi} \cos \theta_i \\
r_{fi} \sin \theta_i \\
t_{fi} \sinh y_{Li} \end{array} \right)
\label{clusterflow}
\end{equation}
Note that $\Sigma$ here is used in a somewhat different way than in the context of hydrodynamics.   In hydrodynamics, $\Sigma^\mu$ is defined as the space-time locus of {\em particle emission}.   In the clustering scenario, it describes the space-time locus of {\em cluster formation}.
Note, in this respect, that $u^{\mu}_i$ and $\Sigma^\mu_i$ are {\em not} fields, but rather arrays of four-vectors, incorporating a finite set of cluster flow velocities and emission coordinates.

Putting everything together, in the lab frame
\begin{equation}
\label{labsource}
S_i (x',p') = \frac{1}{(2\pi)^3}\, \frac{1}{\tau} e^{-E'/T}
e^{-(x^{\prime\alpha} -x_{0i}^\alpha)(x^{\prime}_\beta - x_{0i\beta}) \Lambda^\mu_\alpha
  \Lambda_\mu^\beta /(2 \tau^2)}
\end{equation}
where
\begin{equation}
x_{0i}^{\mu} = \Sigma^{\mu}_i + \tau u^\mu_i 
\end{equation}
and the Lorentz matrix is
\begin{widetext}
\begin{equation}
\Lambda_\nu^\mu = 
\left( 
\begin{array}{cccc}
\gamma_T \cosh y_L & \gamma_T v_T \cosh y_L \cos \theta & \gamma_T v_T
\cosh y_L \sin \theta & \gamma_T \sinh y_L \\
\gamma_T 
v_T \cosh y_L \cos \theta & 1 + \frac{v_T^2 \cos^2
  \theta}{v_T^2+\tanh^2 y_L}\beta &  \frac{v_T^2
  \cos \theta \sin \theta}{v_T^2+\tanh^2 y_L}\beta
&  \frac{v_T \cos \theta \tanh y_L}{v_T^2+\tanh^2 y_L}\beta \\
\gamma_T v_T \cosh y_L  \sin \theta & \frac{v_T^2 \cos \theta \sin
  \theta}{v_T^2+\tanh^2 y_L}\beta &  1+
\frac{v_T^2 \sin^2 \theta}{v_T^2+\tanh^2 y_L}
\beta &  \frac{v_T \sin \theta \tanh y_L}{v_T^2+\tanh^2 y_L}\beta \\
\gamma_T \sinh y_L &  \frac{v_T \cosh \theta \tanh y_L}{v_T^2+\tanh^2
  y_L}\beta &  \frac{v_T \cos \theta
  \tanh y_L}{v_T^2+\tanh^2 y_L}\beta &  1+
\frac{\tanh^2 y_L}{v_T^2+\tanh^2 y_L}\beta \\
\end{array} \right)
\end{equation}
and
\[\ \beta = \gamma_T \cosh y_L - 1  \]
and of course
\begin{equation}
E' = \cosh y_L \gamma_T (E - p_T v_T \cos \left( \Delta \theta \right) - p_L \tanh y_L)
\end{equation}
where $\Delta \theta$ is the relative angle between the direction of the emitted particle and the motion of the cluster.

We note that the emission function of each cluster is in the
Gaussian form
\begin{equation}
S_i (x,p) \sim \exp{\left[-\frac{1}{2\tau^2} (x^\mu - x_0^{\mu})
    B_{\mu}^{ \nu} (x_\nu- x_{0 \nu}) \right]}
\end{equation}
where 
\begin{equation}
B_{\mu}^{ \nu} =  \Lambda_{\mu}^{ \alpha} \Lambda_{\alpha}^{\nu}
\end{equation}
We also need a ``map'' of clusters, giving us the flow and
freeze-out time of cluster $i$.
Assuming boost invariance and ``global'' azimuthal symmetry, as well as small cluster size w.r.t. system size, and even distribution of clusters, we get
\begin{multline}
\sum_l S_i (x,p) =  \sum_{k =1}^{N_r} \sum_{l = 1}^{N_{\theta}^k} \sum_{m=-N_y}^{N_y} S_i \left( u^\mu_i= u^\mu_{nkl},x_{0i} = \Sigma^\mu_{nkl} + \tau u^\mu_{nkl}   \right) \\ 
= \sum_{k =1}^{N_r} \sum_{l =
  1}^{N_{\theta}^k} \sum_{m=-N_y}^{N_y} S_i \left( t_f (\tau_k,y_m),x_f (r_k,\theta_l),y_f (r_k,\theta_l), z_f(\tau_k,y_m),\beta_T (r_k)  \right)
\end{multline}
\end{widetext}
where
\begin{eqnarray}
r_k & = & r_{max} \frac{k}{N_r}\\
\theta_l & = & \frac{2 \pi l}{N_\theta^r}\\
y_m & = & -y_{max}+ 2 m y_{max}
\end{eqnarray}
and the parameters $\beta_T (r_k)$ and $t_f (r_k)$ need to be obtained by ``freezing out'' a hydrodynamic simulation with the appropriate temperature (the temperature, in the QCD phase, where bulk viscosity becomes dominant).

We can estimate the number of clusters in each direction by requiring
each ``side'' of the cluster to be approximately of length $\tau$ (note that
$N_\theta$ depends on $k$)
\begin{eqnarray}
N_r & =  &\frac{r_{max}}{\tau}\\
N_{\theta}^k & = & \frac{2 \pi r_k}{\tau}\\
N_y & = & \tau_k \frac{\sinh (2 y_{max})}{\tau}
\end{eqnarray}
We note that the number of clusters times the number of
particles per cluster is equal to the total multiplicity so
\begin{eqnarray}
N_{cl}
& \sim & r_{max}^2 \tau_{max} \sinh(2 y) \tau^{-3}\\
\ave{N}_i & = & N_{cl}\,  \frac{4 \pi}{(2 \pi)^3} \,  \tau^3 m^2 T K_2 \left( \frac{m}{T} \right) 
\end{eqnarray}
as expected from the statistical model.

To parametrize $t_f,\beta_{T,L} = \tanh (y_{T,L})$ we use the usual
\begin{eqnarray}
t_f & = & \tau_k \cosh y_m\\
x_f & = & r_k \cos (\theta_l)\\
y_f & = & r_k \sin(\theta_l)\\
z_f & = & \tau_k \sinh y_m
\end{eqnarray}
$\beta_T$ and $\tau_k$ can be obtained through a hydrodynamic calculation, assuming freeze-out occurs when $T \sim T_c$.

Throughout, we shall use the usual ``out-side-long'' coordinate
system, where the $x$ axis points
``outwards'', the $y$ axis ``sideways'' and the $z$ axis ``longitudinally'' 

In the mass-shell projection ($k^{\mu} q_{\mu} =0$)
\begin{equation}
q^{\mu} = \left( \begin{array}{c}
\frac{k_o}{k_0} q_o +\frac{k_l}{k_0} q_l \\
q_o \\
q_s\\
q_l \end{array} \right) \qquad  k^{\mu} =
\left(\begin{array}{c} k_0 \\ k_{o} \\ 0 \\ k_l \end{array}  \right)
\end{equation}
\begin{widetext}
\begin{eqnarray}
u^{\mu} k_{\mu} & = & \cosh y_L \cosh y_T k_0 - \sinh y_T \cos \theta
k_o- \sinh y_L \cosh y_T k_L \label{uk} \\
u^\mu q_\mu & = &     \left( \cosh y_L \cosh y_T
\frac{k_o}{k_0} - \sinh y_T \cos \theta
\right) q_o - \sinh y_T \sin \theta q_s   \nonumber \\
& & {} + \left( \cosh y_L \cosh y_T \frac{k_l}{k_0} -
\sinh y_L \cosh y_T\right) q_L  \label{uq}
\end{eqnarray}
and, finally
\begin{eqnarray}
u_\mu x_0^\mu & = & \tau + t_f \cosh y_T - r \sinh y_T \label{x0q}\\
q_\mu x_0^\mu & = & q_\mu \Sigma^\mu+ \tau q_\mu u^\mu  \\
q_\mu \Sigma^\mu & = & \left( \frac{k_o}{k_0} q_o +\frac{k_l}{k_0} q_L \right) t_f
 \cosh(y_L) - r q_o \cos(\theta) - r q_s \sin(\theta) - q_L t_f \sinh(y_L)\, .
\end{eqnarray}


\subsection{A Gaussian approximation estimate}

We use cylindrical symmetry of the emitting function and define $\theta=0$ to be in the
direction of $k$ (the average particle pair momentum).
We immediately remember the standard Gaussian integration formula
\begin{equation}
\label{psi}
\Psi_l (k)= \int d^4 x S_l (x,k) = e^{-u_l^{\mu} k_\mu/T} \frac{(2
  \pi)^2  }{\sqrt{-|B^{\mu \nu}_l|}}
\end{equation}

It is not difficult to prove that
\begin{eqnarray}
\label{expb}
\int x^{\mu} S (x,q) d^4 x & = & \sum_l x_{0l}^\mu \int S_l (x,q) d^4 x =
\sum_l x_{0 l}^{\mu} \Psi_l (k) \\
\int d^4x\, x^{\mu} x^{\nu} S(x,q) & = & \sum_l \left( - 2 \tau^2
\frac{\partial \Psi_l (q)}{\partial B_{l}^{\mu \nu}} + x_{0l}^{\mu}
x_{0l}^{\nu} \Psi_l (k)  \right)\\
-2 \tau^2 \frac{\partial \Psi_l (q)}{\partial B_{l}^{\mu \nu}} & = &
-\tau^2 
\frac{e^{-k_\beta u^\beta_l/T}}{B_{l}^{\mu \nu} \sqrt{ 
-\modulo{B_{l}^{\mu \nu}}}} =-\tau^2 \frac{\Psi_l(k)}{B_{l}^{\mu \nu}}
\end{eqnarray}
Putting everything together we have
\begin{eqnarray}
\ave{\Delta x^\mu \Delta x^\nu}= \frac{\sum_l \left( -\tau^2
  \frac{\Psi_l(k)}{B_{l}^{\mu \nu}} + x_{0l}^\mu x_{0l}^\nu
  \Psi_l(k)\right)}{\sum_l \Psi_l} - \frac{\sum_{lm} x_{0l}^\mu x_{0m}^\nu \Psi_l (k) \Psi_m (k) }{\sum_{lm} \Psi_l (k) \Psi_m (k)}
\end{eqnarray}
which can be used in conjunction with Eqs.~\eqref{rout}, \eqref{rside}, \eqref{rl}
to calculate $R_{o,s,l}$. Note that because of clustering the
emission function {\em can not} be either cylindrically symmetric or boost-invariant, and acquires {\em off-diagonal} terms \cite{kolbangle,wiedemann}:
\begin{eqnarray}
R_{os}^2 = \ave{\Delta x_1 \Delta x_2} - \frac{k_o}{k_0} \ave{\Delta
  x_0 \Delta x_2}\\
R_{sl}^2 = \ave{\Delta x_2 \Delta x_3} - \frac{k_L}{k_0} \ave{\Delta
  x_0 \Delta x_3}
\end{eqnarray}
\begin{eqnarray}
R_{lo}^2 = \ave{\Delta x_0 \Delta x_3} - \frac{k_L}{k_0} \ave{\Delta
  x_1 \Delta x_3}- \frac{k_o}{k_0} \ave{\Delta x_0 \Delta x_1}
+ \frac{k_o k_L}{k_0^2} \ave{\Delta x_0 \Delta x_0}
\end{eqnarray}
while in the cylindrically symmetric case
$R_{os}=R_{sl}=R_{ol}=0$.

Explicitly, the determinant of $B_{\mu \nu}$ is given by \cite{wiedemann}
\begin{eqnarray}
\modulo{ B_{\mu \nu}} =\frac{{\cosh (y_T)}^4
    {\left( -10 - 2\cosh (2y_L) + 
        \cosh (2\left( y_L - 
            y_T \right) ) + 
        2\cosh (2y_T) + 
        \cosh (2\left( y_L + 
            y_T \right) ) \right) }^2}{64} \label{detB}
\end{eqnarray}
and 
\begin{eqnarray}
B_{0 0} & = &
{\cosh (y_T)}^2\,
  \left( {\cosh (y_L)}^2\,
     {\cosh (y_T)}^2 + 
    {\sinh (y_L)}^2 \right)
\\
B_{0 1} &= &
2\,\cos (\theta)\,{\cosh (y_L)}^2\,
  {\cosh (y_T)}^2\,\sinh (y_T)
\\
B_{1 1} & = &
{\cos (\theta)}^2\,{\cosh (y_L)}^2\,
   {\cosh (y_T)}^2\,
   {\sinh (y_T)}^2 + 
  \frac{{\cos (\theta)}^2\,{\cosh (y_L)}^2\,
     {\sinh (y_L)}^2\,
     {\sinh (y_T)}^2}{{\left( 1 + 
        \cosh (y_L)\,\cosh (y_T)
        \right) }^2} +
\nonumber \\
&& +
  \frac{{\cos (\theta)}^2\,{\cosh (y_L)}^4\,
     {\sin (\theta)}^2\,{\sinh (y_T)}^4}
     {{\left( 1 + \cosh (y_L)\,
         \cosh (y_T) \right) }^2} + 
  {\left( 1 + \frac{{\cos (\theta)}^2\,
         \left( -1 + \cosh (y_L)\,
            \cosh (y_T) \right) \,
         {\sinh (y_T)}^2}{{\sinh (
            y_T)}^2 + 
         {\tanh (y_L)}^2} \right) }^2 
\\
B_{2 2} & =  &
{\cosh (y_L)}^2\,{\cosh (y_T)}^2\,
   {\sin (\theta)}^2\,{\sinh (y_T)}^2 + 
  \frac{{\cosh (y_L)}^2\,{\sin (\theta)}^2\,
     {\sinh (y_L)}^2\,
     {\sinh (y_T)}^2}{{\left( 1 + 
        \cosh (y_L)\,\cosh (y_T)
        \right) }^2} + 
\nonumber \\
&& 
+ \frac{{\cos (\theta)}^2\,{\cosh (y_L)}^4\,
     {\sin (\theta)}^2\,{\sinh (y_T)}^4}
     {{\left( 1 + \cosh (y_L)\,
         \cosh (y_T) \right) }^2} + 
  {\left( 1 + \frac{\left( -1 + 
           \cosh (y_L)\,
            \cosh (y_T) \right) \,
         {\sin (\theta)}^2\,{\sinh (y_T)}^2}
         {{\sinh (y_T)}^2 + 
         {\tanh (y_L)}^2} \right) }^2
\\
B_{33} & = & \frac{2 + 6\,\cosh (2\,y_L) + 
    \cosh (2\,\left( y_L - 
        y_T \right) ) - 
    2\,\cosh (2\,y_T) + 
    \cosh (2\,\left( y_L + 
        y_T \right) )}{8}
\\
B_{02} & = & 2\,{\cosh (y_L)}^2\,
  {\cosh (y_T)}^2\,\sin (\theta)\,
  \sinh (y_T)
\\
B_{03} & = & {\cosh (y_T)}^2\,\sinh (2\,y_L)
\\
B_{12} & = &
\frac{{\cosh (y_L)}^2\,
    \left( 3 + \cosh (2\,y_T) \right) \,
    \sin (2\,\theta)\,{\sinh (y_T)}^2}{4}
\\
B_{13} & = & \frac{\cos (\theta)\,\left( 3 + 
      \cosh (2\,y_T) \right) \,
    \sinh (2\,y_L)\,\sinh (y_T)}{
    4}
\\
B_{23} & = & \frac{\sin (\theta)\,\left( 3 + 
      \cosh (2\,y_T) \right) \,
    \sinh (2\,y_L)\,\sinh (y_T)}{
    4} \, .
\end{eqnarray}

\subsection{Calculation of the full correlation function}

The full correlation function is given by \cite{wiedemann}
\begin{equation}
C(k,q) = 1 + \frac{\modulo{ \sum_i \tilde{S}_i (k,q)}^2}{
\left[ \sum_i \Psi_i \left( k-\frac{1}{2} q \right) \right]\left[
  \sum_j \Psi_j \left( k+\frac{1}{2} q \right) \right] }
\end{equation}
We can again use the standard formulae regarding Fourier transforms of
Gaussians, where
\begin{equation}
\tilde{S}_l (q,p) =  \Psi(k_\mu) e^{i q_\alpha x_{0 l}^{\alpha}} G_l (q^\mu)
=  e^{-k_{\mu} u_l^\mu/T}  \frac{(2 \pi)^2}{\sqrt{-|B_{\mu
 \nu}^l|}} e^{i q_\alpha x_{0l}^{\alpha}} G_l (q^\mu)
\end{equation}
where
\begin{equation}
\label{gaussfour}
G(q^\mu) = \exp \left[-  \frac{\tau^2}{2} q^{\mu}
 (B_{\mu \nu}^l)^{-1} q^{\nu}  \right]\, .
\end{equation}
Note that, as a result of interference {\em between} clusters, the
correlation coefficient will pick up an interference pattern (which
will disappear if the spatial size $\tau$ varies considerably from
cluster to cluster).
The FT of the whole system is, up to a factor of $2 \tau (2 \pi)^4$
(that cancels),
\begin{eqnarray}
\modulo{ \sum_l \tilde{S}_l (k,q)}^2 = 
 \sum_l \frac{e^{-2 u_l^\mu k_\mu/T}}{|B_{\mu \nu}^l|} G_l^2 (q^{\mu})
  + \sum_{l \ne m}  \frac{e^{-((u_l^\mu +u_m^\mu) k_\mu /T}}{\sqrt{|B_{\mu \nu}^l|
 |B_{\mu \nu}^m|}} \cos \left( q_\mu (x_{0l}^\mu - x_{0m}^\mu) \right)
  G_l (q^\mu) G_m (q^\mu) 
\end{eqnarray}
the bottom of the correlation is, up to the same factor,
\begin{eqnarray}
\left[ \sum_i \Psi_i \left( k-\frac{1}{2} q \right) \right] \left[ 
  \sum_j \Psi_j \left( k+\frac{1}{2} q \right) \right] = \sum_l \frac{e^{-2 u_l^\mu k_\mu/T}}{|B_{\mu \nu}^l|}
+ \sum_{l \ne m}  \frac{ e^{-(u_l^\mu + u_m^\mu)k_\mu} \cosh\left(
  \frac{(u_l^\mu - u_m^\mu) q_\mu }{2 T}  \right)}{\sqrt{|B_{\mu \nu}^l|
 |B_{\mu \nu}^m|}}
\end{eqnarray}
\end{widetext}
If clusters all have the same size, the cosine term would give a characteristic oscillating pattern.   However, for a general cluster distribution such terms should in general interfere destructively.   
Still, the form of the tail should be considerably different from the Gaussian 
approximation, especially in the high $q_{o,l}$ tail.

The explicit form of the exponent in the Fourier transform,  
$q^{\mu} (B_{\mu \nu}^l)^{-1} q^{\nu}$ , which goes into Eq.~\eqref{gaussfour} 
is of course a closed formula, but too long to be included here.

Generalizing this formalism to non-central, non boost-invariant collisions is straight-forward.  
It is also possible to generalize this approach to a distribution of cluster sizes, by updating 
Eq.~\eqref{comovingsource} with $\tau \rightarrow \tau_i$ and Eq.~\eqref{clustersum} with 
$\sum S_i \rightarrow \sum f(\tau_i) S_i$, where $f(\tau_i)$ is the cluster probability distribution.
For realistic cluster distributions, however, Monte Carlo methods might prove necessary.


\end{document}